\documentclass{aa}
\usepackage{natbib}
\bibpunct{(}{)}{;}{a}{}{,}
\usepackage{graphicx}
\begin{document}
\title{XMM-Newton observation of the bursting\\ pulsar GRO J1744-28 in quiescence}
\author{Fr\'{e}d\'{e}ric Daigne$^{1}$, Paolo Goldoni$^{1}$, Philippe Ferrando$^{1}$, Andrea Goldwurm$^{1}$, Anne Decourchelle$^{1}$ and Robert S. Warwick$^{2}$}
\institute{CEA/DSM/DAPNIA, Service d'Astrophysique, C.E. Saclay, 91191 Gif sur Yvette Cedex, France
\and Department of Physics and Astronomy, University of Leicester, Leicester LE1 7RH, United Kingdom}
\mail{daigne@discovery.saclay.cea.fr}
\date{Received 12.12.01 / Accepted 11.02.02}
\abstract{The XMM-Newton X-ray observatory performed a pointed observation of
 the bursting pulsar GRO J1744-28 in April 2001 for about 10 ks during a program devoted to the scan of the Galactic center region. 
After the discovery of this source by BATSE in December 1995 during a very active bursting phase, it has been in quiescence since April 1997. We present here the first detection of GRO J1744-28 in its quiescent state at a position which is consistent with previous high-energy positions (but not consistent with a proposed IR counterpart). The observed luminosity of the source in quiescence is about 6 orders of magnitude weaker than the luminosity in outburst.
\keywords{Pulsars: individual: GRO J1744-28; Stars: neutron; X-rays: bursts}}
\titlerunning{XMM observation of GRO J1744-28 in quiescence}
\authorrunning{Daigne et al.}
\maketitle

\section{Introduction}
The bursting pulsar GRO J1744-28 is a unique source that presents both the characteristics of a type II X-ray burster and a X-ray pulsar. It
was discovered near the Galactic center by the Burst and Transient Source Experiment (BATSE) on board the Compton Gamma-Ray Observatory (CGRO) on 1995 December 2 \citep{fishman:95,kouveliotou:96a}, when it was experiencing a very active bursting phase with a rate of about 20 hard X-ray bursts per hour. After one day, the source entered a regime of hourly bursting which lasted until 1996 April 26. Many of these bursts were also detected by the Proportional Counter Array (PCA) instrument on board the Rossi X-ray Timing Explorer (RXTE) \citep{swank:96,giles:96} and the Oriented Scintillation Spectrometer Experiment(OSSE) instrument on board CGRO \citep{strickman:96}. Then the burst rate decreased rapidly and the bursting activity ceased for the next 7 months. On 1996 December 1, a second bursting phase started 
with a similar burst rate to that seen in the first outburst \citep{woods:99}.
 It lasted until 1997 April 7. The source was no longer detected by BATSE or RXTE after this date. More than 5800 bursts were observed by BATSE in total.\\
During the first outburst, 
persistent hard X-ray emission 
from GRO J1744-28 was detected by
BATSE and OSSE \citep{paciesas:96}. \citet{finger:96} measured that GRO J1744-28 is a 467 ms X-ray pulsar in a binary system of orbital period 11.8 days. The pulsar period determination was confirmed with ASCA 
\citep{dotani:96a,dotani:96b}. It has also been found that the X-ray bursts were showing pulsations with the same 467 ms period \citep{kouveliotou:96b} and had similar hard X-ray / soft gamma-ray spectra than the persistent emission \citep{briggs:96,strickman:96}, so that the association is firmly established.\\
The initial source localization was a $6^\circ$ radius error circle \citep{fishman:95}, which was rapidly refined to a 24' wide annulus by the Interplanetary Network using time-delay measurements between CGRO and other satellites, mainly Ulysses \citep{hurley:95}. Observations with ASCA provided a $\sim 1'$ error circle. 
ROSAT \citep{kouveliotou:96c,augusteijn:97}
improved the source localization to R.A. $17^\mathrm{h}44^\mathrm{m}33^\mathrm{s}.1$ and decl. $-28^{\circ}44'29''$ (J2000) with a $\sim 10''$ error circle. An optical / near-infrared variable object was observed in 1996 at the periphery of this ROSAT error circle \citep{augusteijn:97,cole:97} and proposed as a possible counterpart.
More recently, \citet{hurley:00} derived a more precise IPN localization of GRO J1744-28 with a $3\sigma$ error ellipse of $532\ \mathrm{arcsec^{2}}$ fully contained in the ASCA error circle and partially overlapping with the ROSAT error circle. The optical / near-infrared variable source lies outside the new IPN $3 \sigma$ ellipse, which makes it unlikely to be associated with GRO J1744-28. \\
\citet{finger:96} determined the mass function of the binary system to be $1.36\ 10^{-4}\ \mathrm{M_{\odot}}$. This supports that
GRO J1744-28
is
very likely a Low-Mass X-ray Binary (LMXB), where the central neutron star accretes matter through Roche lobe overflow from an evolved low-mass stellar companion \citep{daumerie:96,sturner:96}.  
This is however the only known example of such a system showing both the characteristics of a X-ray pulsar (the persistent pulsed emission) and of a type II X-ray burster. The relative fluences in the bursts and in the persistent emission probably rule out a thermonuclear origin of the bursts, which are more likely of type II. Such bursts are accretion powered and are possibly triggered by an instability in the accretion flow, the nature of which is not clearly identified \citep{kouveliotou:96b,lewin:96,cannizzo:96}. This is supported by several similarities between the bursts of GRO J1744-28 and those of the Rapid Burster MXB 1730-355 \citep{lewin:96}, i.e the first and unique object where type II bursts were discovered.\\
In this paper, we report new observations of GRO J1744-28 in the 
X-ray band, made with XMM-Newton in April 2001 as it was in its quiescent state. 
\begin{table}[!t]
\resizebox{\hsize}{!}{
\begin{tabular}{lccccc}
\hline
Inst. & RA & DEC & stat. & error & error\\
      &    &     & error & (RA)  & (DEC)\\
\hline
\hline
PN   & $17^\mathrm{h}43^\mathrm{m}51^\mathrm{s}.3$ & $-28^{\circ}46'40''$ & $\sim 0.6''$ & $-0.3''$ & $+1.8''$\\
MOS1 & $17^\mathrm{h}43^\mathrm{m}51^\mathrm{s}.4$ & $-28^{\circ}46'38''$ & $\sim 0.7''$ & $-1.2''$ & $+0.1''$\\
MOS2 & $17^\mathrm{h}41^\mathrm{m}51^\mathrm{s}.4$ & $-28^{\circ}46'37''$ & $\sim 0.7''$ & $-1.7''$ & $-0.8''$\\
\hline
\end{tabular}
}
\caption{Position of the reference star TYC 6840-38-1 as detected by each of the EPIC instruments. The first three columns (right ascension, declination and statistical error at $1 \sigma$ level) are the result of the standard SAS source detection procedure. The last two columns show the 
measured offsets with respect to the Tycho-2 catalogue position
R.A. 
$17^\mathrm{h}43^\mathrm{m}51^\mathrm{s}.3$
and decl. 
$-28^{\circ}46'38''$.}
\label{tab:star}
\end{table}

\section{Observations and data analysis}
\label{sec:analysis}
\begin{figure}
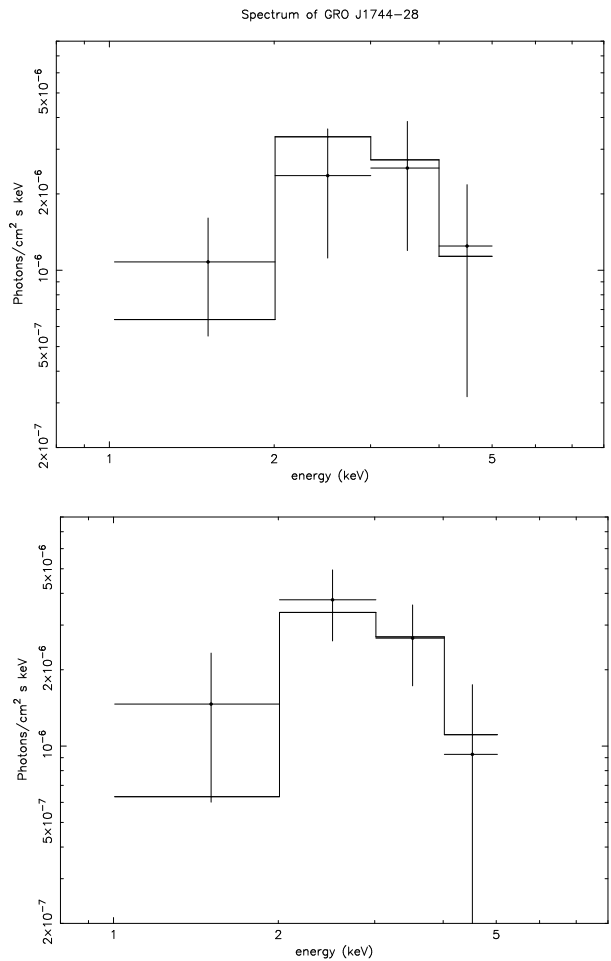

\begin{center}
\includegraphics[angle=-90,width=0.9\hsize]{h3472f1a.eps}
\vspace*{3ex}

\includegraphics[angle=-90,width=0.9\hsize]{h3472f1b.eps}
\end{center}
\caption{Spectrum of GRO J1744-28 : the flux is plotted as a function of energy for the PN camera (top) and the sum of the MOS1 and MOS2 cameras (bottom). The solid line corresponds to the fit of the data by an absorbed blackbody spectrum with a column density $5.1\ 10^{-22}\ \mathrm{cm^{-2}}$ and a temperature $kT = 0.6\ \mathrm{keV}$.}
\label{fig:spectrum16}
\end{figure}

GRO J1744-28 was observed with the European Photon Imaging Camera (EPIC), both with the MOS cameras \citep{turner:01}
on 2001 April 4 at UT 13:14:27 and with the
PN camera \citep{struder:01}
on the same day starting at UT 14:27:23. 
The MOS and PN cameras
were operating in full window mode with the medium filter. Data reduction was performed using the XMM SAS (Science Analysis Software) and the standard XSPEC packages.\\
The time histogram of all events above 10 keV showed 
that the 
background has only small variations,
without any very intense flare, even if the averaged flux is close to the alert level as defined in the standard SAS flare rejection procedure.
For this reason we applied no time filtering so that the total duration has been kept to $9199\ \mathrm{s}$ for the MOS 1 \& 2 cameras and $4699\ \mathrm{s}$ for the PN camera. 
A preliminary visual inspection of the MOS 1,2 and PN images showed that a faint source was present at the position of GRO J1744-28 (with no significant flux below 1 keV and above 6 keV). The absence of flux below 1 keV can be explained by the strong absorption towards the Galactic center.
Therefore all the further analysis has been made in the 1-6 keV band. Also, it could be noticed that for some unidentified reasons, the background in the MOS2 image appears to be higher than in  the MOS1 image.\\
In this energy band, we generated an image for each of the three EPIC instruments, keeping only the valid X-ray events using the canonical range of pattern specified by the SAS (0-12 for the MOS cameras and 0-4 for the PN camera). We applied to these three images the standard source detection procedure with the following SAS tasks : (i) \textit{eexpmap} and \textit{emask} to generate the exposure map and the detection mask for each image; (ii) \textit{eboxdetect} (in local mode) to generate an input list of source positions; (iii) \textit{esplinemap} to generate the corresponding background map; (iv) \textit{eboxdetect} (in map mode) to generate a new list of sources with a better detection sensitivity
using the background map; (v) \textit{emldetect} to perform a maximum likelihood PSF fit to the count distribution of each source. This procedure was also applied to an image made of the sum of all X-ray events of the three instruments in the 1-6 keV band.\\
\begin{table}[!t]
\begin{center}
\begin{tabular}{lcccc}
\hline
Inst. & RA & DEC & stat.  & counts\\
      &    &     & error  & \\
\hline
\hline
PN   & $17^\mathrm{h}44^\mathrm{m}33^\mathrm{s}.1$  & $-28^{\circ}44'26''$ & $\sim 2.4''$ & $22.4 \pm 6.1$\\
MOS1 & $17^\mathrm{h}44^\mathrm{m}33^\mathrm{s}.2$  & $-28^{\circ}44'27''$ & $\sim 1.9''$ & $27.4 \pm 6.0$\\
MOS2 & $17^\mathrm{h}44^\mathrm{m}33^\mathrm{s}.1$  & $-28^{\circ}44'16''$ & $\sim 6.0''$ & $18.8 \pm 7.4$\\
EPIC & $17^\mathrm{h}44^\mathrm{m}33^\mathrm{s}.2$  & $-28^{\circ}44'26''$ & $\sim 1.4''$ & $59.2 \pm 10.0$\\
\hline
\end{tabular}
\end{center}
\label{tab:position}
\caption{Position of GRO J1744-28 as detected by each of the EPIC instruments. The right ascension (RA), declination (DEC), statistical error at $1\sigma$ level and counts number are the result of the SAS standard source detection procedure.
The last line (EPIC) corresponds to the image made of the sum of the PN, MOS1 and MOS2 images.}
\end{table}
The standard SAS source detection procedure that we applied to the data found one weak source within the ROSAT $\sim 10''$ error circle of GRO J1744-28. This source was also present in the standard SAS pipeline output.
As can be seen in Table \ref{tab:position}, it is a detection at the $\sim 4\sigma$ level for the PN and MOS1 cameras and a detection at the $2.5\sigma$ level only for the MOS2 camera.
The number of sources at the $3\sigma$ level in the total field of view of the MOS cameras (radius $15'$) is about 100 so that the probability to find such a source by chance in an area comparable with the ROSAT error box (radius $\sim 10''$) is about 1 \%
.
Then we identify this weak source with GRO J1744-28.\\
The position error obtained with the standard SAS source detection procedure can be dominated by the errors of the absolute astrometry (maximum $\sim 4''$). To estimate these systematic errors,
we proceeded to identify strong sources in our field of view. Unfortunately only one is strong enough to be used in this way. This bright source which was detected by each instrument has been identified -- based on positional coincidence --
as the star TYC 6840-38-1 in the Tycho-2 Catalogue \citep{hog:00} which contains the position, the proper motion and the two-colour photometric data for the 2.5 million brightest stars in the sky. Its catalogue position is R.A. $17^{h}43^{m}51.^{s}3$ and decl. $-28^{\circ}46'38''$ (with a negligible proper motion for our purpose).
Table \ref{tab:star} shows the position of this star obtained with each instrument. They are all consistent with the catalogue position, taking into account the statistical error given by the \textit{emldetect} procedure ($\sim 0.7''$) and the quoted accuracy of the XMM-Newton coordinates\,: the MOS1 and MOS2 positions are distant of $\sim 1''$ from each other whereas the PN position is shifted of $\sim 3''$ with respect to the MOS2 position.
This is in agreement with the negligible offset found in the MOS 1\&2 roll angle and the mean offset of $\sim -0.30^{\circ}$ found in the PN roll angle \citep{tedds:01}.
As the star is located at $\sim 10'$ of the center of the field of view, the expected shift in the PN position is indeed $\sim 3''$. However the same shift at the position of GRO J1744-28 is negligible because the source is at the center of the field of view. The nice agreement bewteen the position of the star obtained with the three EPIC instruments and the catalogue position shows that for our observation, the systematic error in the position is $\sim 1''$.\\ 
The position of GRO J1744-28 is the same in the three instruments taking into account the statistical error (see Table \ref{tab:position}). The best accuracy is obtained when summing the events of the three instruments. The derived position is R.A. $17^{h}44^{m}33^{s}.2$ and decl. $-28^{\circ}44'25''$ 
 with an error circle of about $4''$ at the $3 \sigma$ level (summing in quadrature a statistical error of $\sim 4''$ and a systematic error $\sim 1''$).
The
closest other source 
is located at $\sim 25''$ of GRO J1744-28, well outside the ROSAT error circle, and therefore can be easily rejected.\\
To construct the spectral distribution of the photons,
we have decided to sum the two MOS cameras events 
to increase the statistics, which is possible because the response matrix are similar.
We extracted from the PN and MOS 1+2 images the X-ray events 
within a radius of $15''$ of the GRO J1744-28 position.
We grouped the 
photons in 5 energy bands\,: $1-2\ \mathrm{keV}$, $2-3\ \mathrm{keV}$, $3-4\ \mathrm{keV}$, $4-5\ \mathrm{keV}$ and $5-6\ \mathrm{keV}$. Background was estimated using offset regions in the same central CCD. 
After substracting it, we obtained a detectable flux only in the $1-5$ keV band.
The resulting spectrum was obtained with XSPEC and is plotted in Fig.~\ref{fig:spectrum16}. The source count rate in the 1-6 keV energy band is $\left(4.75 \pm 1.32\right)\ 10^{-3}\ \mathrm{counts/s}$ for the PN camera and $\left(3.40\pm 0.76\right)\ 10^{-4}\ \mathrm{counts/s}$ for the MOS 1+2. Fixing the hydrogen column density to the value of $5.1\times 10^{22}\ \mathrm{cm^{-2}}$ \citep{dotani:96a}, we tried to fit the spectrum of GRO J1744-28 either with a power-law or with a blackbody distribution. 
We found that acceptable fits of the observed spectrum  with a power-law could be obtained for a photon index in the range $\sim 2-5$. Acceptable fits with a blackbody distribution could be obtained as well for a temperature $kT$ in the range $\sim 0.4-1.\ \mathrm{keV}$.
The corresponding unabsorbed energy flux in the $1-5\ \mathrm{keV}$ is about $2-5\ 10^{-13}\ \mathrm{erg.cm^{-2}.s^{-1}}$ in the power-law case and about $1-2\ 10^{-13}\ \mathrm{erg.cm^{-2}.s^{-1}}$ in the blackbody case.
\section{Results and discussion}
\subsection{Source position}
\begin{figure}
\includegraphics[width=\hsize]{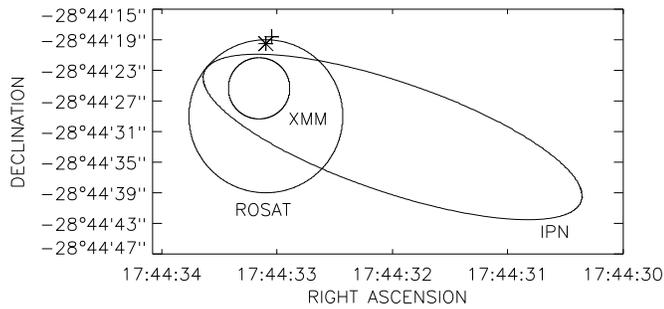}
\caption{An $\sim 1' \times 1'$ square region containing the approximate $3\sigma$ IPN error ellipse, the $\sim 10''$ radius ROSAT error circle and the $3\sigma$ XMM error circle ($\sim 4''$ radius). The two positions of the proposed optical / near-infrared counterpart \citep{augusteijn:97,cole:97} are respectively marked with a star and a cross.}
\label{fig:position}
\end{figure}

Fig.~\ref{fig:position} shows the XMM-Newton error circle we have derived in the last section. It is entirely included in the intersection of the ROSAT error circle \citep{augusteijn:97} and the IPN error ellipse \citep{hurley:00}. This strengthens the identification of the source observed by XMM-Newton with GRO J1744-28, despite the fact that this source is too weak to allow 
the search for pulsations (such pulsations were also impossible to study with the ROSAT observation). 
The association of 
the proposed optical / near-infrared counterpart \citep{augusteijn:97,cole:97}
with GRO J1744-28 
is rejected without any ambiguity with
the more precise XMM-Newton error circle we have obtained in the previous section.
It is then important to carry out deeper searches for the real IR counterpart of GRO J1744-28 using the better determined XMM-Newton position.

\subsection{Source luminosity in quiescence}
Despite a very low statistics, we tried in the last section to estimate the spectrum of GRO J1744-28 in quiescence. This spectrum appears to decrease very fast at high energy, as it is indicated by the photon index found when fitting with a power-law.
The unabsorbed luminosity in the 1-5 keV band is
\begin{equation}
L \simeq \left( 1 - 6 \right) 10^{33}\ d_{10}^{-2}\ \mathrm{erg/s}
\end{equation}
where $d_{10}$ is the distance of the source in unit of 10 kpc. The quiescent luminosity is then about five orders of magnitude smaller than the outburst luminosity 
\begin{equation}
L \simeq 10^{38}\ d_{10}^{-2}\ \mathrm{erg/s}\ 
\end{equation}
and is compatible with the typical quiescent luminosity of a neutron star transient \citep{asai:98}. On the other hand, if we assume a blackbody distribution, the range in temperature we obtain (0.4-1. keV) is slightly outside the typical range of neutron star transients in quiescence ($kT \sim 0.2-0.3$ keV; \citet{asai:98}). The temperature we find corresponds to an emitting area 
\begin{equation}
A = \frac{L}{\sigma T^{4}} \simeq \left( 10^{9} - 10^{11} \right)\ d_{10}^{-2}\ \mathrm{cm^{2}}\ .
\end{equation}
Such an area corresponds to a sphere of radius $10^{4}--10^{5}\ \mathrm{cm}$ which is well below the size of a neutron star. Then the observed luminosity cannot be associated with the thermal emission of the total surface of a non accreting neutron star. Many other origins can be proposed, such as the emission due to the accretion on the magnetic poles or the emission from the accretion onto the neutron star magnetosphere \citep{campana:98}. However
the large uncertainty we get on the spectral distribution of the X-ray photons makes very difficult any detailed comparison with a particular model.

\begin{acknowledgements}
F.D. acknowledges financial support from a postdoctoral fellowship from the French Spatial Agency (CNES).
\end{acknowledgements}

\bibliographystyle{aa}
\bibliography{h3472}
\end{document}